# Deadline-Chasing in Digital Health: Modeling EMR Adoption Dynamics and Regulatory Impact in Indonesian Primary Care

Suryo Satrio
Faculty of Computer Science
Universitas Indonesia
Jakarta, Indonesia
suryo.satrio91@ui.ac.id

Bukhori Muhammad Aqid
Klinik Pintar
Jakarta, Indonesia
aqid@klinikpintar.id

*Abstract—* Indonesia's digital healthcare transformation is accelerating under Minister of Health Regulation No. 24/2022, which mandates the adoption of Electronic Medical Records (EMR) and integration with the SATUSEHAT platform. However, empirical evidence regarding the factors, trajectory and speed of adoption in Primary Health Facilities (FKTP) remains limited. This study aims to evaluate the level and rate of EMR adoption within the customer network of a major EMR system provider (PT MTK) and model short-term projections. This is an observational study with the main variables being cumulative registered EMR facilities, monthly registration flow, same-month activation, same-month inactivation, and the estimated number of eligible FKTPs nationally/monthly (eligible facilities). The analysis uses descriptive analysis, logistic growth modeling, and ARIMA forecasting. The results of the study over 33 months showed that cumulative registered facilities increased from 2 to 3,533, with a median same-month activation rate of 0.889 (IQR 0.717–0.992). The proportion of final adoption compared to eligible facilities was 8.9% (3,533 of 39,852). The ARIMA model projects a cumulative ~3,997 clinics (95% CI 3,697–4,298) by June 2025. The estimated growth in logistics converges with a carrying capacity of ~4.1 thousand facilities. The study findings reveal that EMR adoption within the customer network of EMR system providers is showing steady growth with rapid activation in the month of registration. Although the cumulative series showed no major departures from the long-term trend, localized step-ups around deadlines suggest "deadline-chasing," so impact should be maximized by aligning interventions to the deadline calendar. Given the trajectory, total market share of FKTP for PT MTK remains <10% at the end of 2024, but continues to increase in 2025.

## I. INTRODUCTION

Digitization of primary care plays a critical role in the quality, safety, and efficiency of healthcare services. Electronic Medical Records (EMR) are associated with improved service coordination, documentation quality, and population data utilization, although barriers to adoption remain (cost, workload, interoperability) [1, 2]. The Indonesian government mandated the implementation of EHRs through Minister of Health Regulation No. 24 of 2022 (Permenkes No. 24/2022) and established the SATUSEHAT ecosystem for data integration based on Health Level Seven International (HL7) Fast Healthcare Interoperability Resources (FHIR) [4, 15].

Indonesia's Minister of Health (Kemenkes) Regulation No. 24/2022 mandates EMR adoption and SATUSEHAT integration by December 31, 2023 to drive digital transformation. Noncompliance carries administrative sanctions, from written warnings to recommendations for accreditation adjustment or revocation via the Director General [3]. A 2023 circular sets phased enforcement: through December 31, 2023 written warnings for lacking integrated EMR, and by March 31, 2024 accreditation adjustments if EMR is implemented but not integrated. By July 31, 2024, facilities transmitting <50% visit data to SATUSEHAT face accreditation adjustments, with revocation recommended for those not implementing EMR or out-of-building service recording at all. By December 31, 2024, Community Health Centers with <100% data submission remain subject to accreditation adjustments, and the Minister may also request business-permit revocation.

Recognizing the scale of transformation and integration, the Ministry of Communication and Informatics (Kominfo), through the Digital Economy Directorate, collaborated with the Indonesian Healthtech Association to initiate an EMR adoption initiative, partnering with PT MTK as one of the Electronic System Providers (PSE), to identify barriers and formulate implementation solutions in healthcare facilities, particularly clinics and independent practices. This initiative was realized through a series of activities throughout 2023, including March 18, May 6, June 17, and September 9 [9].

This study analyzes operational data at PT. Medigo Teknologi Kesehatan (MTK), an EMR provider that integrates the registration-to-payment flow, to: (1) measure adoption rates and dynamics, including activation by registration month; (2) model short-term growth and projections. We hypothesize a consistent growth trend and high initial activation, but moderate penetration across the entire primary care population.

## II. LITERATURE REVIEW

### A. Primary Health Care Facilities

Primary Health Care Facilities (FKTP) serve as the frontline of Indonesia's healthcare system, serving as the first point of contact for communities to access promotive, preventive, curative, and rehabilitative services [3]. FKTPs include Community Health Centers (Puskesmas), primary care clinics, general practitioners' practices, dental practices, and equivalent private healthcare facilities.

### B. Electronic Medical Record

Electronic Medical Records (EMR) are digital systems for recording patient health information, replacing traditional paper-based medical records [3]. EMR contains comprehensive data such as patient identity, diagnoses, treatments, and medical procedures, which can be securely accessed by authorized healthcare professionals. In Indonesia, the adoption of EMR is regulated by Minister of Health Regulation No. 24/2022, which requires all healthcare facilities to implement EMR and integrate them with the SATUSEHAT platform.

### C. Digital System Adoption Factors

EMR adoption is multifactorial, shaped by individual, organizational, technological, and environmental determinants. At the user level, perceived usefulness and ease of use together with performance/effort expectations, facility conditions, and social influence drive intention to use [5, 6]. At organizational and ecosystem levels, infrastructure readiness, leadership support, training and resources, vendor support, external pressures, and regulations are pivotal, while reviews consistently flag workflow fit, interoperability, system quality/usability, costs and incentives, privacy/security, and implementation/post-implementation support as key barriers or enablers [7, 8]. At the governance level, interoperability standards and strong privacy/security compliance underpin scalability, trusted data exchange, and sustainable adoption [2].

Table I. Adoption of Clinical Digitalization in the Kominfo Workshop Process Event Participants

| Process stages | Bogor | Banyumas | Purbalingga |
|---|---|---|---|
| Clinic Contacted | 350 | 400 | 400 |
| Event Activation | 71 | 100 | 100 |
| Agree to Adopt EMR | 68 | 85 | 111 |
| EMR Training | 44 | 37 | 102 |
| Trial Stage | 32 | 43 | 90 |
| Active Usage | 25 | 25 | 76 |

A Kominfo-linked survey of 1,150 FKTP engaging with PT MTK across Bogor, Banyumas, and Purbalingga identified major concerns shaping EMR adoption (Table I). Top issues included medical data/transaction security (≈65–79%), data migration (≈62–86%), and post-implementation technical support (≈57–63%) (Table II). Additional inhibitors were being in an initial/trial phase (50% Bogor; 77.2% Banyumas), small patient volumes (10.7% Bogor), and limited human resources (3.5%

Bogor; 20% Purbalingga), underscoring the need for targeted risk-mitigation and mentoring (Table III). Feature priorities in Purbalingga centered on BPJS bridging (PCare/JKN Mobile, 87.5%), core EMR modules (78.5%), online registration/queuing (64.29%), with several others at 50-57.14% [9]. Collectively, these concerns and feature needs map closely to daily clinical workflows and financing, streamline accreditation and SATUSEHAT integration, reduce administrative burden, and improve patient experience, making them prerequisites for sustainable EMR adoption in primary care.

Table II. Clinician Concerns in EMR Adoption

| Factors | Bogor | Banyumas | Purbalingga |
|---|---|---|---|
| Medical data and transaction security | 64.63% | 66% | 78.57% |
| Data transfer/migration (patient data, medical records, etc.) | 62.20% | 61.7% | 85.71% |
| Support for handling technical issues after implementation | 57.32% | 62.8% | - |
| Investment in IT personnel to manage the system | 54.88% | - | 62.49% |
| System training and change management | 51.22% | - | - |
| Completeness of features/models according to needs | 46.34% | - | - |
| Digital system investment | 43.90% | - | - |

Table III. Factors Inhibiting EMR Adoption

| Factors | Bogor | Banyumas | Purbalingga |
|---|---|---|---|
| Still in the initial/trial stage | 50% | 77.2% | - |
| Patient volume (feeling there are few patients) | 10.7% | - | - |
| Other systems already in place | 17.8% | - | 20% |
| Human resource constraints | 3.5% | - | 20% |
| Infrastructure/equipment constraints | 3.5% | - | 26.6% |
| Other constraints (unwillingness to adopt digital services, etc.) | 14.5% | 22.8% | 23.4% |

D. Logistics Growth Model

The Logistic Growth Model is a mathematical model that describes population growth, which initially increases rapidly, then slows down and eventually reaches a plateau [10]. Mathematically, the logistic growth model is written as $\frac{dN}{dT} = rN(1 - \frac{N}{K})$, where $N$ is the population size, $r$ is the intrinsic growth rate, and $K$ is the environmental carrying capacity. When the population is still far below the carrying capacity, growth approaches exponential. However, as the population approaches $K$, the factor $(1 - \frac{N}{K})$ causes the growth rate to decrease until it eventually reaches equilibrium. In the realm of innovation diffusion, the work of Bass (1969) and the review by Meade & Islam describe that growth models (logistic, Gompertz) are commonly used for forecasting product/technology penetration [11, 12].

E. ARIMA

ARIMA (Autoregressive Integrated Moving Average) is a time-series modeling approach used to analyze and forecast sequential data. It combines autoregressive (AR) terms linking current to past values, an integration (I) step that differs the series to achieve stationarity, and moving-average (MA) terms that model serial correlation in past forecast errors. An ARIMA (p, d, q) model specifies the AR order p, differencing degree d, and MA order q; typical workflow includes identification via autocorrelation (ACF)/partial autocorrelation (PACF), parameter estimation, and performance evaluation to reduce prediction error. ARIMA is widely applied across domains such as economics, finance, and operations for capturing short-term dynamics and forecasting from historical data.

### III. METHODOLOGY

The method used in this study is a retrospective longitudinal observational study with a descriptive-analytical quantitative approach and case study method.

A. Design

An observational time-series study (Apr 2022–Dec 2024) using aggregated operational data from customers of PT MTK. Reporting follows relevant STROBE (observational) and TRIPOD principles (for modeling/forecasting transparency).

B. Background and Participants

The unit of analysis was primary healthcare facilities (clinics, independent practices) that registered/used the PT MTK EMR system during the observation period. Hospitals were not included.

C. Intervention/Technology

The EMR system is an integrated end-to-end solution from registration to payment (registration, clinical records, pharmacy, billing) with integration support to SATUSEHAT (HL7 FHIR) for data exchange.

D. Data Sources and Variables

This study used monthly, facility-level aggregates from a company database with no patient identifiers. Missing values were set to zero only for rate calculations when denominators existed, with no carry-forward between months, and sensitivity checks compared inflows derived from cumulative differences versus explicit monthly columns. The dataset contained no Protected Health Information, and per institutional policy, the work qualified as non-human subjects research requiring no formal ethics approval and involved no clinical intervention. The EMR platform operates under ISO 27001–aligned controls (access control, audit trails, encryption in transit/at rest), and potentially riskier linkages were not enabled. Potential selection bias (single-vendor network) and measurement bias (definition of "active") were mitigated through explicit operational definitions and sensitivity analyses. The following are variable definitions based on classification:
- month: observation month (YYYY-MM).
- eligible_facilities: number of national FKTP per month (market space indicator).
- cumulative_reg: accumulation of facilities that have registered/created an EMR account.
- monthly_inflow_explicit: clinics registered that month (active + inactive).
- total_active: clinics that register and become active in the same month.
- inactive: clinics that registered but were not active in the same month.
- cumulative_active: accumulation of previously active accounts.

From this variable it is then derived: same-month activation rate = total_active / monthly_inflow_explicit, adoption vs eligible = cumulative_reg / eligible_facilities.

E. Analysis

Descriptive analysis includes cumulative trends, inflows, and same-month activation rates (median, IQR).

A logistic growth model ( $y\_t = K/(1 + A e^{(-r t)})$ ) was fitted to cumulative_reg ($t = 0, ..., T − 1$) using nonlinear least squares with non-negative bounds. Initial values: ( $K = 1.2 * max(y)$ ), ( $A = (K − y\_0)/y\_0$ ) when (y_0>0), (r=0.2). Parameter confidence intervals (CIs) were calculated from the asymptotic covariances if convergence occurred.

The best model for ARIMA forecasting is ARIMA (2,1,2) and was selected based on the Akaike Information Criterion (AIC) of cumulative_reg. The 6-month projection and 95% IK are calculated from the point forecast (predicted mean) and conf_int.

### IV. DISCUSSION

A. Result

The dataset spans 33 months (April 2022–December 2024). Cumulative

registered facilities increased from 2 (starting) to 3,533 (ending). Eligible facilities increased from 29,656 to 39,852 over the observation period.

The monthly registration flow (active + inactive) yielded a median same-month activation rate of 0.889 (IQR 0.717–0.992), indicating that the majority of registered facilities activated their accounts within the same month. Compared with the estimated eligible facilities, the cumulative adoption rate by December 2024 was 8.9% (3,533/39,852).

Table IV. Sample summary and adoption metrics

| Start month | End month | Months observed | Eligible facilities (first) | Eligible facilities (last) | Cumulative registered (start) | Cumulative registered (end) | Same-month activation rate (median) | Same-month activation rate (IQR) |
|---|---|---|---|---|---|---|---|---|
| 2022-04 | 2024-12 | 33 | 29656 | 39852 | 2 | 3533 | 0.889 | (0.717, 0.992) |

Figure I. Cumulative trajectory of EMR adoption, logistic fit , and 6-month ARIMA projections with 95% CI

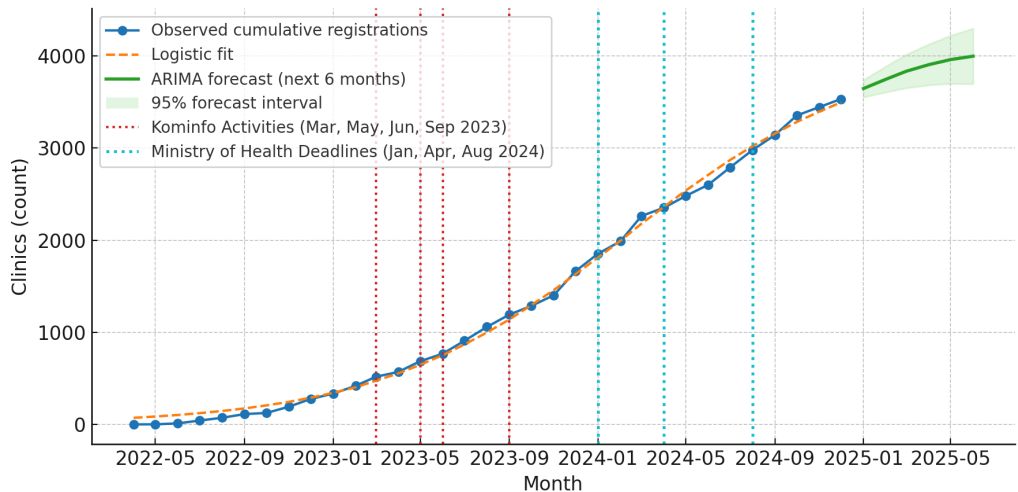

Parameter estimates converged with a carrying capacity (K) of 4,112 facilities and a rate (r) of 0.18 per month. This indicates a flattening growth curve, consistent with the intermediate diffusion phase in the primary health care market.

The 6-month projection ARIMA indicates cumulative registered facilities reaching ~3,997 (95% CI 3,697–4,298) by June 2025, representing an absolute increase of ~464 facilities from December 2024 (Figure 1, attached projection table). This forecast is conservative because it is based on a cumulative series, partially damping cyclical variations in monthly flows.

There is no System Usability Scale (SUS) survey in this dataset; however, the high same-month activation rate serves as an indicator of early implementation and onboarding readiness.

B. Main Findings

EMR adoption on the PT MTK network increased from 2 to 3,533 facilities in 33 months, with median activation in the month of registration at ~89%, a strong indicator of successful onboarding and operational readiness. Penetration of the FKTP remained <10% at the end of 2024, indicating substantial room for growth. Logistics models place capacity saturation at approximately 4,100 facilities on the current trajectory, while ARIMA forecasting indicates ~3,997 facilities by June 2025. The study findings align with international literature demonstrating the benefits of EMR as well as adoption challenges (cost, interoperability, workload) [8]. Policy and interoperability support (SATUSEHAT, HL7 FHIR) facilitates EMR diffusion through data exchange standards and integration governance.

The high level of activation during the registration month also indicates vendor support and adequate organizational facilitation conditions aligned with performance/effort expectancy determinants. Survey feedback that BPJS bridging features (PCare/JKN Mobile), core EMR modules, online registration and queuing, inventory and finance, reporting, billing payments, and private insurance integration are "must-have" requirements, explains the high initial conversion rate. These features directly impact clinical processes and the financing ecosystem, reducing administrative burdens and accelerating SATUSEHAT integration compliance [9, 15]. Conversely, concerns about security and privacy, data migration, IT HR readiness, and training/post-implementation needs have the potential to hold back expansion from ~9% to the next diffusion phase. Mitigation measures require packages such as a standard migration toolkit, technical support SLAs, and a structured change management program.

Based on the Kominfo's activities (March, May, June, and September 2023 vertical markers in Figure 1), a visual inspection of the cumulative curve indicates a trajectory that remains consistent with logistics growth, with indications of a temporary acceleration around Q2/2023. Because the cumulative curve dampens fluctuations in inflows, this study does not infer causality. However, conceptually, socialization/mentoring activities can act as normative pressure and salience cues that increase adoption intentions in facilities already considering adoption, especially when combined with the implementation package (short training, help desk, and FHIR integration playbook).

The time series findings of 8.9% cumulative penetration in December 2024, with a median same-month activation of 0.889, align well with the barriers highlighted in the survey (Table III). One factor, namely, the initial/trial stage, indicates that many primary health care facilities are still in the exploration phase, leading to delayed adoption decisions. Barriers include low patient volumes that weaken the perceived return on investment, limited human resources and infrastructure/equipment factors that reduce operational readiness, existing systems that create migration friction and potential vendor lock-in, and other barriers, such as inability to be contacted/reluctance to adopt, reflecting socio-organizational barriers. The pattern in Figure 1 (with the 2023 Kominfo Activity marker) indicates that outreach campaigns have the potential to trigger episodic spikes in already "warm" groups, but do not automatically penetrate segments with structural barriers. Therefore, the effectiveness of Kominfo's activities will likely increase significantly if they are packaged with interventions that directly target dominant barriers per region, such as micro-training and multi-layered helpdesks for human resource shortages, device/connectivity support for infrastructure, and volume-proportional incentive/fee schemes for low-patient clinics, so that public communication initiatives are connected to real technical and organizational readiness.

Ministry of Health Deadlines (January, April, and August 2024), policy markers such as deadlines often serve as coercive pressures that drive compliance near the deadline. In the cumulative series of this study, no major deviations from the long-term trajectory were observed, but localized step-ups around the deadline period are consistent with "deadline-chasing" behavior. Based on these results, for a productive effect (rather than simply enrollment without activation), it is recommended that operational interventions be implemented and aligned with the deadline calendar, a countdown campaign 4–6 weeks in advance, fast-track onboarding (including a secure data migration package), a clinic success kit (SOPs, micro-training materials), and clearly communicated security/compliance assurances. Recommended metrics for monitoring the impact of deadlines are same-month activation per cohort of enrollees, conversion of inactive to active within ≤60 days, and usage stability (≥X encounters/month) to ensure that the surge near the deadline truly translates into meaningful and sustained usage.

C. Limitations

Internal validity is limited by operational, aggregate data and a system-defined "active" status (activation in the registration month), which does not capture longer-term use.

External validity is constrained by a single-vendor sample, limited survey data, and possible measurement bias from national "eligible facilities" estimates that may shift over time. Modeling limits include logistic fit sensitivity to the observation window and conservative ARIMA forecasts on cumulative series that omit external drivers (e.g., policy incentives, market changes).

D. Future Research

To rigorously test whether outreach/mentoring acts as normative pressure and salience cues, use an interrupted time series on monthly registrations with pre-trend tests (placebo leads), seasonal controls, and robust standard errors. Future work should incorporate usage intensity and data-quality metrics, usability (SUS) and workflow-fit assessments, and subgroup (clinic type, location) plus economic (cost/ROI) analyses. Additionally, evaluate FHIR interoperability and data-quality rules to ensure complete and accurate SATUSEHAT submissions.

V. CONCLUSION AND IMPLICATIONS

Adoption of EMR in primary care within the network of EMR providers, in this study, PT MTK, has shown steady growth with high initial activation, but deployment relative to all primary health care providers nationwide remains moderate. With regulatory support and interoperability, near-term projections indicate continued improvement. A focus on onboarding, interoperability, and organizational change support will accelerate clinically valuable diffusion.

There are four implications for practice and operations, particularly for EMR service providers. First is to focus on rapid activation, High same-month activation demonstrates the impact of streamlined onboarding and implementation support. Second is scalability, to drive penetration above 10–15%, a scale-out strategy needs to prioritize areas with high primary health care (FKTP) density and digital readiness. Third is operational interventions need to be implemented and aligned with policymakers' deadlines to encourage high system activation. Fourth is SATUSEHAT Integration, FHIR integration compliance (e.g., SNOMED CT terminology) accelerates cross-system value-add.